\documentclass[11pt,a4paper]{article}
\usepackage[T1]{fontenc}
\usepackage[utf8]{inputenc}
\usepackage{authblk}
\usepackage{algorithm}
\usepackage{algorithmic}
\usepackage{paralist}
\usepackage{amsmath,amssymb}
\usepackage{fancyhdr}
\usepackage{hyperref}
\usepackage{pgf}
\usepackage{tikz} \usetikzlibrary{automata, arrows}

\usepackage{pgfplots}
\pgfplotsset{compat=newest}
\usepgfplotslibrary{units}

\usetikzlibrary{arrows,positioning,calc}
\usepackage{tikz}
\usepackage{tkz-graph}
\usetikzlibrary{backgrounds}
\usepgflibrary{shapes}
\usetikzlibrary{through}
\usepgflibrary{shapes.symbols} 
\usepgflibrary[shapes.symbols] 
\usepgflibrary{shapes.symbols}
\usetikzlibrary{shapes.symbols} 
\usetikzlibrary[shapes.symbols] 
\usepgflibrary{shapes.geometric} 
\usepgflibrary[shapes.geometric] 
\usetikzlibrary{shapes.geometric} 
\usetikzlibrary[shapes.geometric] 
\usetikzlibrary{automata,positioning}
\usetikzlibrary{shapes,snakes}

\newcommand{\dx}[1]{\,\mathrm{d} #1}
\newcommand{\fd}[1]{\mathrm{D}_*^{\alpha}\, #1}
\newcommand{\ceil}[1]{\left\lceil #1 \right\rceil}

\tikzset{
    >=stealth',
    smallpunkt/.style={
               rectangle,
               rounded corners,
               draw=black, very thick,
               text width=1.5em,
               minimum height=2em,
               text centered},
    punkt/.style={
           rectangle,
           rounded corners,
           draw=black, very thick,
           text width=6.5em,
           minimum height=2em,
           text centered},
     largepunkt/.style={
                rectangle,
                rounded corners,
                draw=black, very thick,
                text width=9.5em,
                minimum height=2em,
                text centered},
    pil/.style={
           ->,
           thick,
           }
}

\pgfdeclarelayer{background}
\pgfdeclarelayer{foreground}
\pgfsetlayers{background,main,foreground}

\usepackage{epsfig}
\usepackage{amsthm}

\def\squarebox#1{\hbox to #1{\hfill\vbox to #1{\vfill}}}

\newcommand{\dahntab}[1]{
  \newbox\mybok%
  \setbox\mybok=\hbox{\vbox{
      \begin{tabbing}
        #1
      \end{tabbing}%
    }}

  \newdimen\bokwidth%
  \bokwidth=\wd\mybok%
  \newdimen\myl%
  \myl=\textwidth%
  \divide\myl by 2%
  \divide\bokwidth by -2%
  \advance\myl by\bokwidth%
  \vrule width\myl height 0pt depth 0pt%
  \usebox\mybok%
}

\begin{document}

\title{HPC optimal parallel communication algorithm for the simulation of fractional-order systems}
\author{Cosmin BONCHI\c{S}}
\author{Eva KASLIK}
\author{Florin RO\c{S}U}
\affil{West University of Timi\c{s}oara and the eAustria Research Institute, Bd. V. P\^arvan 4, cam 045B, Timi\c{s}oara, RO-300223, Romania.}

\def\titlerunning{HPC optimal parallel communication algorithm for the simulation of fractional-order systems}
\def\authorrunning{C. BONCHI\c{S}, E. KASLIK \& F. RO\C{S}U}

\maketitle

\begin{abstract}
A parallel numerical simulation algorithm is presented for fractional-order systems involving Caputo-type derivatives, based on the Adams-Bashforth-Moulton (ABM) predictor-corrector scheme. The parallel algorithm is implemented using several different approaches: a pure MPI version, a combination of MPI with OpenMP optimization and a memory saving speedup approach. All tests run on a BlueGene/P cluster, and comparative improvement results for the running time are provided. As an applied experiment, the solutions of a fractional-order version of a system describing a forced series LCR circuit are numerically computed, depicting cascades of period-doubling bifurcations which lead to the onset of chaotic behavior.  
\end{abstract}


\section{Introduction}

Compared to their integer-order counterparts, over the past decades, fractional-order dynamical systems have proved to provide more accurate and realistic results in the modeling of real world processes arising from diverse applied fields \cite{Cottone}.

Although many qualitative properties of fractional-order systems can be studied by analytical tools (such as local stability of equilibrium states), theoretical characterization of chaos in fractional- order dynamical systems is yet to be investigated. In order to assess chaotic behavior of fractional order dynamical systems, accurate estimation of the solutions over large time intervals is of utmost importance. However, an essential observation is that the employed discretization should use a small step size, with the aim of providing an accurate estimation to the solution of the fractional-order system under investigation.  

Several numerical methods are used for fractional-order systems, such as generalizations of predictor-corrector methods \cite{daftardar2014new,Diethelm,garrappa2010linear}, p-fractional linear multi-step methods \cite{Galeone_2009,pedas2011spline} or the Adomian decomposition method \cite{daftardar2005adomian,duan2012review,song2013new}. These numerical schemes have a major drawback due to the non-locality of the fractional differential operators which reflects the hereditary nature of the problem: in order to obtain a reliable estimation of the solution, at every iteration step, all previous iterations have to be taken into account. Therefore, this implies extreme computational costs whenever the solution is computed over a large time interval, with a small step size. While the numerical computation of a solution of an ordinary fractional differential equation on a fixed interval $[0, T]$, by one of the standard algorithms described above, has an arithmetic complexity
of $O(h^{-2})$ (where $h$ denotes the step size), in the case of ordinary differential equations of first order, the arithmetic complexity is only $O(h^{-1})$ \cite{baleanu2016fractional}. Several approaches have been used to deal with these difficulties, such as the short memory principle \cite{deng2007short,deng2012numerical} or the nested mesh scheme \cite{ford2001numerical}. However, a loss of accuracy is inevitable for both methods, mainly due to the fact
that parts of the integration interval are simply ignored.

Nevertheless, these difficulties may be overcome using parallel computing algorithms implemented in a conventional way or using available high performance computing systems \cite{Baban2016,Cafagna-1}. 

In this paper, we will present an efficient parallel algorithm implemented using Message Passing Interface (MPI) and running on a high performance computing system BlueGene/P cluster that has 1024 processors and 4TB of RAM memory. The numerical method considered here for implementing the fractional-order system is the Adams-Bashforth-Moulton predictor-corrector scheme \cite{Diethelm}. The main challenge for implementing this method is to parallelize the computation of the solution because, the computation of an iteration step requires to take into account all previous iterations.

\section{Preliminaries}
Consider an ordinary fractional differential equation of the form:
\begin{equation}\label{eq:fde}
\begin{cases}
\fd{y(t)} = f(t, y(t)), & t \in [0, T] \\
y^{(k)}(0) = y_0^{k}, & k \in \{0, \dots, \ceil{\alpha} - 1\},
\end{cases}
\end{equation}
where $\alpha>0$ and $\ceil{\cdot}$ denotes the ceiling function that rounds up to the nearest integer. The fractional derivative of Caputo-type is defined as:
\[
\fd{y(t)} = \frac{1}{\Gamma(\ceil{\alpha} - \alpha)}
            \int_0^T \frac{y^{(\ceil{\alpha})}(t)}{(t - \tau)^{\alpha - \ceil{\alpha}+ 1}} \dx{\tau}.
\]

The numerical method used in this paper to solve \ref{eq:fde} is a fractional version of the Adams-Bashforth-Moulton predictor corrector scheme \cite{Diethelm}. The domain $[0, T]$ is discretized into $N$ intervals with a step size $h = \frac{T}{N}$ and the grid points $t_n = nh$, for $n \in \{0, \dots, N\}$. We will also denote $y_n = y(t_n)$ and $f_n = f(t_n, y_n)$ with $y_0=y_0^{0}$ as the initial condition.

The first step of the scheme is the \textbf{predictor}, which will give a
first approximation $y_{n + 1}^P$ of our solution:
\begin{equation}\label{eq:abm}
y_{n + 1}^P = \sum_{k = 0}^{\ceil{\alpha} - 1} \frac{t_{n + 1}^k}{k!} y_0^{(k)} +
              h^\alpha \sum_{k = 0}^n b_{n - k} f_k,
\quad\text{where}~
b_n = \frac{(n + 1)^\alpha + n^\alpha}{\Gamma(\alpha + 1)}.
\end{equation}
The final approximation of the solution, called the \textbf{corrector}, is given by:
\[
y_{n + 1} = \sum_{k = 0}^{\ceil{\alpha} - 1} \frac{t_{n + 1}^k}{k!} y_0^{(k)} +
            h^\alpha \left(c_n f_0 + \sum_{k = 1}^n a_{n - k} f_k +
                           \frac{f(t_{n + 1}, y^P_{n + 1})}{\Gamma(\alpha + 2)}
                     \right),
\]
where the weights $a_n$ and $c_n$ are defined as:
\[
a_n = \frac{(n + 2)^{\alpha + 1} - 2(n + 1)^{\alpha + 1} + n^{\alpha + 1}}{\Gamma(\alpha + 2)}
\quad\text{and}\quad c_n = \frac{n^{\alpha + 1} - (n - \alpha)(n + 1)^\alpha}{\Gamma(\alpha + 2)}.
\]
This numerical scheme can be generalized in a straight-forward way, when one has to deal with a system of fractional-order differential equations. 

The main computational difficulty of this scheme arises from the fact that at each step, we require the complete history of the variable, i.e., when computing $y_{n + 1}$, we need to know all previous values $y_k$ that are used to compute $f_k$, for $k \leq n$. This makes numerical methods addressed at solving fractional differential equations (or systems) notoriously hard to parallelize.

\section{Parallel numerical algorithm}

The parallel implementation of Adams-Bashforth-Moulton algorithm as a  method for solving a fractional-order dynamical system was first presented by Diethelm \cite{Diethelm_parallel}. The solution presented there is not suitable to be running on a HPC cluster because there is an unbalanced workload and waiting (idle) times for the processes that can cause the performance to be very low, due to the HPC parallel implementation rules \cite{HPCDev}. Also, the amount of messages passed between processes does not respect the HPC parallel implementation idea \cite{HPCDev}, and strongly influence the overall performance.

\subsection{The classical parallel approach}

In the previous works \cite{Baban2016,RosuCMMSE}, we explored how numerical implementations of the Adams-Bashforth-Moulton method for fractional-order systems can be accelerated by using parallel computing techniques. We investigated the feasibility of parallel computing algorithms and their efficiency in reducing the computational costs over a large time interval. The results in \cite{Baban2016} concerning the parallel implementation for the Adams-Bashforth-Moulton method on HPC and CUDA show that some execution times are quite high and thus, they limit the time frame needed for more accurate simulations.

\begin{figure}[h!]
	\centering	
	\includegraphics [width=0.55 \textwidth]{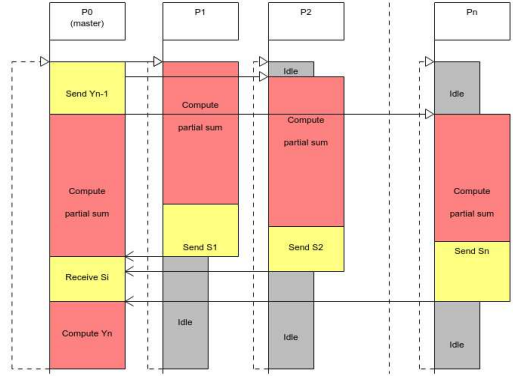}
	\centering	
	\caption{One master process}
	\label{fig:hpcclassic}       
\end{figure}

In \cite{Baban2016}, the HPC implementation was made using a classical approach: a process is the master and all others are slaves to help compute the values for the predictor and corrector. The classical execution flow is presented in Figure \ref{fig:hpcclassic} and shows that the master process is working either by computing or communicating, while the slave processes have idle times. On a HPC architecture these idle times could causes huge delays in communication and drastically increase the overall simulation time. Due to this low performance, the research of optimizing the HPC solution was further pursued, generating the results presented in \cite{RosuCMMSE}.

\subsection{Parallel implementation using pure MPI}

In order to improve the overall simulation time, the workload has been improved in \cite{RosuCMMSE}. First, the idle times of the processes have been removed. Secondly, we decreased the number of messages, which implied saving times in the communication part.

\begin{figure*}[h!]
\centering
  \includegraphics [width=0.65 \textwidth]{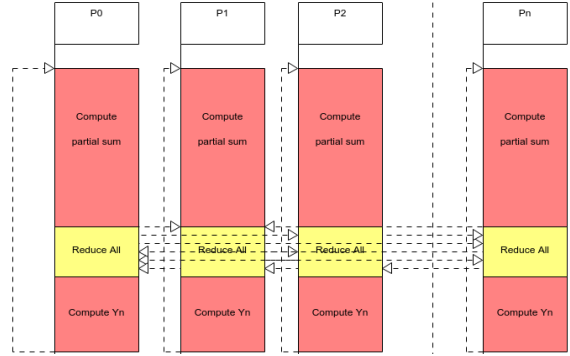}
\caption{No master process}
\label{fig:hpcreduce}       
\end{figure*}

\begin{algorithm}[h!]
    \caption{Parallel Algorithm for the Adams-Bashforth-Moulton scheme.}
    \label{alg}
    {$T$ end of the time interval.} \\
    {$N$ global number of points.}  \\
    {$P$ number of processes.}   \\
    {$p$ current process.}  \\
    $N_P \leftarrow N / P$\; \\
    $y_0 \leftarrow$ initial condition\; \\
    {\bf for} $n \in [1, N]$ \\
    $~~~~~~   Sp \leftarrow 0$\; \\
    $~~~~~~   Sc \leftarrow 0$\; \\
    $~~~~~~   n_{min} \leftarrow N_Pp$\; \\
    $~~~~~~   n_{max} \leftarrow N_P(p + 1)$\; \\
    $~~~~~$  \textit{compute local sum for predictor and corrector} \\
    $~~~~~$ {\bf for} $k \in [n_{min}, n_{max}]$ \\
    $~~~~~~~~~~~~   Sp \leftarrow Sp + b_{n - k} f_k$\; \\
    $~~~~~~~~~~~~   Sc \leftarrow Sc + c_{n - k} f_k$\;  \\
    $~~~~~$  \textit{compute the global sum and sent to all processes}\\
    $~~~~~$  \texttt{MPI\_Allreduce}$(Sp,Sc)$\;\\
    $~~~~~$  \textit{compute the predictor at time $t_n$}\\
    $~~~~~~   y_n^P \leftarrow y_0 + a_n * Sp$\;\\
    $~~~~~~   $\textit{compute the corrector at time $t_n$}\\
    $~~~~~~   y_n \leftarrow y_0 + c_n * (y_n^P + Sc)$\; 
\end{algorithm}

The parallel implementation method presented in Algorithm \ref{alg} reflects the core of our implementation. The computation for the partial sum is done by each process and the final sum for the predictor and corrector is reduced to all processes (by \texttt{MPI\_Allreduce}$(Sp,Sc)$). So instead of having the classical architecture where one process is the master and all the other processes are slaves just to compute the partial sum, in this approach all processes $P$ compute the iteration function values ($y_n$),  and all act as master processes. 

In our approach (designed in Figure \ref{fig:hpcreduce}) we have been able to avoid idle times for the processes, obtaining a more balanced work load. Another advantage is that the overall communication between processes was reduced by removing the messages between the slave and master processes. There is only one message being exchanged when the global sum is reduced to all processes, and because the workload inside the process is balanced, the synchronized exchange time is very short.
A similar method/solution was also presented and tested on a PC using a CPU core by \cite{XingCai} with very interesting results from the parallel computational point of view.

An in-depth analysis of the architecture of the available BlueGene/P cluster (used for simulations) reveals the advantages provided by the hardware capabilities which further reduce the communication between the processes. Our actual BlueGene/P cluster consists of 1024 nodes,  one node having 4G bytes of RAM and a quad-core processor. 

In order to efficiently use the resources, the processes are launched in Virtual Node (VN) mode accordingly to \cite{HPCDev}. In VN mode each process is executed by only one core from an available node. Hence, on the BlueGene/P cluster, four processes are executed on each physical node. In this way, a maximal number of 4048 parallel processes can be executed. On our tests we run 1024 processes in VN mode, and thus we actually use 256 physical nodes.

\subsection{Optimal communication time using MPI and OpenMP}

Aiming for a more efficient use of the BlueGene resources, we combine the MPI with OpenMP capabilities in our new implementation. This leads to a full employment of the 256 physical nodes as follows.

In order to run one process in one physical node we change the previous Virtual Node mode by Symmetrical Multiprocessing mode. Thus, a process can use all the cores from the processor, and the computation can be parallelized by multi-threading. Therefore, by using 256 processes we obtain 1024 computational threads but the MPI messages are exchanged only between 256 instead of 1024 processes.

Using OpenMP, the improvement of the previous Algorithm \ref{alg} is reflected in Algorithm \ref{algOpenMP} at line 8, which computes the partial sum by using all local cores. 

\begin{algorithm}[h!]
    \caption{Parallel Algorithm using OpenMP}
    \label{algOpenMP}
    {$T$ end of the time interval.} \\
    {$N$ global number of points.}  \\
    {$P$ number of processes.}   \\
    {$p$ current process.}  \\
    $N_P \leftarrow N / P$\; \\
    $y_0 \leftarrow$ initial condition\; \\
    {\bf for} $n \in [1, N]$ \\
    $~~~~~~   Sp \leftarrow 0$\; \\
    $~~~~~~   Sc \leftarrow 0$\; \\
    $~~~~~~   n_{min} \leftarrow N_Pp$\; \\
    $~~~~~~   n_{max} \leftarrow N_P(p + 1)$\; \\
    $~~~~~$  \textit{compute local sum for predictor and corrector} \\
    $~~~~~$        \textit{\texttt{\#pragma omp parallel for reduction(+:$Sp,Sc$)}}\; \\
    $~~~~~$ {\bf for} $k \in [n_{min}, n_{max}]$ \\
    $~~~~~~~~~~~~   Sp \leftarrow Sp + b_{n - k} f_k$\; \\
    $~~~~~~~~~~~~   Sc \leftarrow Sc + c_{n - k} f_k$\;  \\
    $~~~~~$  \textit{compute the global sum and sent to all processes}\\
    $~~~~~$  \texttt{MPI\_Allreduce}$(Sp,Sc)$\;\\
    $~~~~~$  \textit{compute the predictor at time $t_n$}\\
    $~~~~~~   y_n^P \leftarrow y_0 + a_n * Sp$\;\\
    $~~~~~~   $\textit{compute the corrector at time $t_n$}\\
    $~~~~~~   y_n \leftarrow y_0 + c_n * (y_n^P + Sc)$\; 
\end{algorithm}

On our BlueGene/P cluster, the processor in one node is quad-core. Thus, we can use four parallel threads to compute the partial sum. Therefore, only 256 processes are used instead of 1024 as in the pure MPI implementation, the communication between processes being much faster while the computational time is the same.

\subsection{Memory saving improves computing time}
All previous tests were done using \textit{long double} data types, because the simulations need high precision and they are very costly in CPU operations and memory usage. 

The \textit{long double} data is classically represented in the memory on 10 bytes with the precision of 21 decimal points. However, a closer look at the memory usage (in the current hardware) by changing data type to \textit{double} has a major impact on computation time, due to data alignment. Having the data aligned by 32 bytes, the data access is faster, and improves the overall performance by a factor of at least 10 compared to the MPI and OpenMP implementations.

The drawback is a decrease in precision. For 3 millions steps, the precision is still conserved by $10^{-6}$, so it can be used for an overall view of the evolution of the numerical solution, and then, for a more precise simulation, \textit{long double} version can be used.
Even with this drawback, the performance in computation time is a good compromise, as it can be seen in the simulation results. 

\section{Simulation results}
We implemented and tested the presented approach using the HPC cluster of the West University of Timi\c{s}oara (Romania), namely, a BlueGene/P cluster that consists of a fully loaded single BlueGene/P rack that has 1024 quad-core CPUs and 4TB of RAM memory and can offer a performance up to 11.7 TFlops.

\begin{table}[htbp]
\caption{Simulation results in seconds for different numbers of time steps} 
\label{table:times}
\centering 
\begin{tabular}{c c c c c}
\hline\hline                        
\#steps  & HPC classic & pure MPI & MPI and OpenMP &  Memory saving
\\  [0.5ex] 
\hline\hline                 
1000000 & 4621.25 & 549.64 & 506.58 & 69.06 \\
1500000 & 9162.33 & 1158.13 & 1059.35 & 115.67 \\
2000000 & 14931.16 & 2009.87 & 1810.72 & 169.23 \\
2500000 & 22697.66 & 3066.92 & 2762.92 & 234.56\\
3000000 & 31659.66 & 4381.42 & 3912.29 & 304.17 \\ 
[1ex]      
\hline 
\end{tabular}
\end{table}

In Table \ref{table:times} we present the simulation run time results (in seconds) using different number of time steps (number of global points). The total running times of the HPC Classic approach have been obtained by the algorithm presented in \cite{Baban2016}. With the HPC pure MPI implementation by the algorithm presented in \cite{RosuCMMSE}, the running time decreased by a factor of 8 with respect to classical approach \cite{Baban2016}. Although in these two approaches the pure computation time is similar, we emphasize that the overall running time is massively improved, due to the synchronized communication. 

Moreover, using the MPI combined with OpenMP approach, we can see an improvement of around 10\% at the overall running time, compared to the pure MPI version. This is due to the fact that using the same resources, instead of having 1024 processes running, we only have 256 processes that have to communicate between each other. We emphasize that the computation time is the same, and the improvement is due to the communication time.

\begin{figure}[htbp]
\begin{center}

\includegraphics[width=0.8 \textwidth]{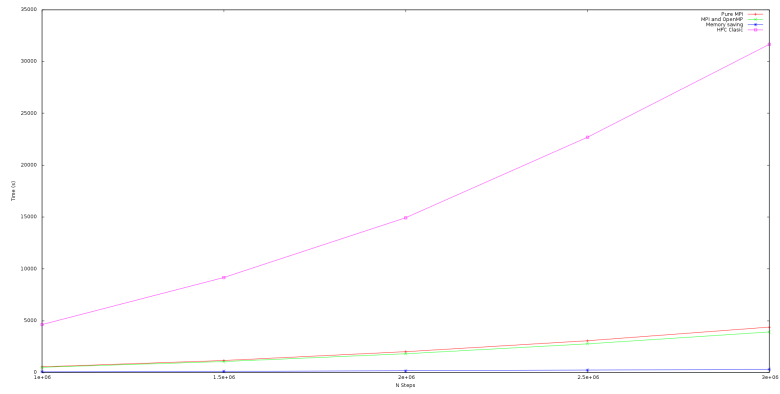}
\caption{Running time comparison for all approaches}
\label{fig:hpccompareall}
\end{center}
\end{figure}

The graphical representation of the running times (Figure \ref{fig:hpccompareall}) clearly presents the speed up using our different approaches of parallelizing the algorithm. Studying the available hardware architecture, we were able to improve the communication time, confirmed by the 10\% efficiency obtained in practice.

\begin{figure}[htbp]
	\begin{center}
    \includegraphics[width=0.8 \textwidth]{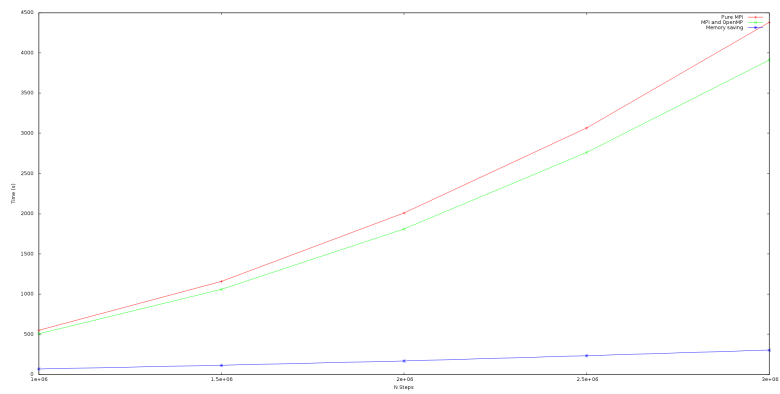}
		\caption{Running time comparison for optimized approaches}
 \label{fig:hpccompare}
 \end{center}
 \end{figure}

Paying attention to data usage, the computation time is massively improved, having the simulations running very fast. The execution time (see Figure \ref{fig:hpccompare}) is improved by a factor of at least 10 compared to the execution running time in the MPI and OpenMP implementations.

The non-classical parallel approaches are fast enough to enable us to run simulations with even more than 5 million steps. These improvements allow us to compute the numerical solution of a fractional-order system over a large number of steps, providing a better understanding about the system's behavior from the dynamic point of view. The next section includes more details about the numerical analysis of a test system.

\section{Numerical experiment}

Our test case is the fractional-order version of the normalized system describing a forced series LCR circuit \cite{Palanivel2017}:
\begin{equation}\label{sys.LRC}
\begin{cases}
 D^{\alpha_1} x(t) =y-g(x)\\
 D^{\alpha_2} y(t) =-\sigma y-x+f\sin(\omega t)
\end{cases}
\end{equation}
where $\alpha_1,\alpha_2\in(0,1)$, $\sigma,f,\omega>0$ and the function $g$ is piecewise linear and is defined as: 
\[g(x)=
\begin{cases}
bx-a+b, & \text{if } x\leq -1\\
ax,  & \text{if } |x|<1\\
bx+a-b, & \text{if } x\geq 1
\end{cases}
\]
The parameter values considered for the numerical simulations are: $\sigma=1.015$, $\omega=0.55$, $a=-1.02$ and $b=-0.58$. 

\begin{figure}[h!]
\begin{center}
\includegraphics*[width=\textwidth]{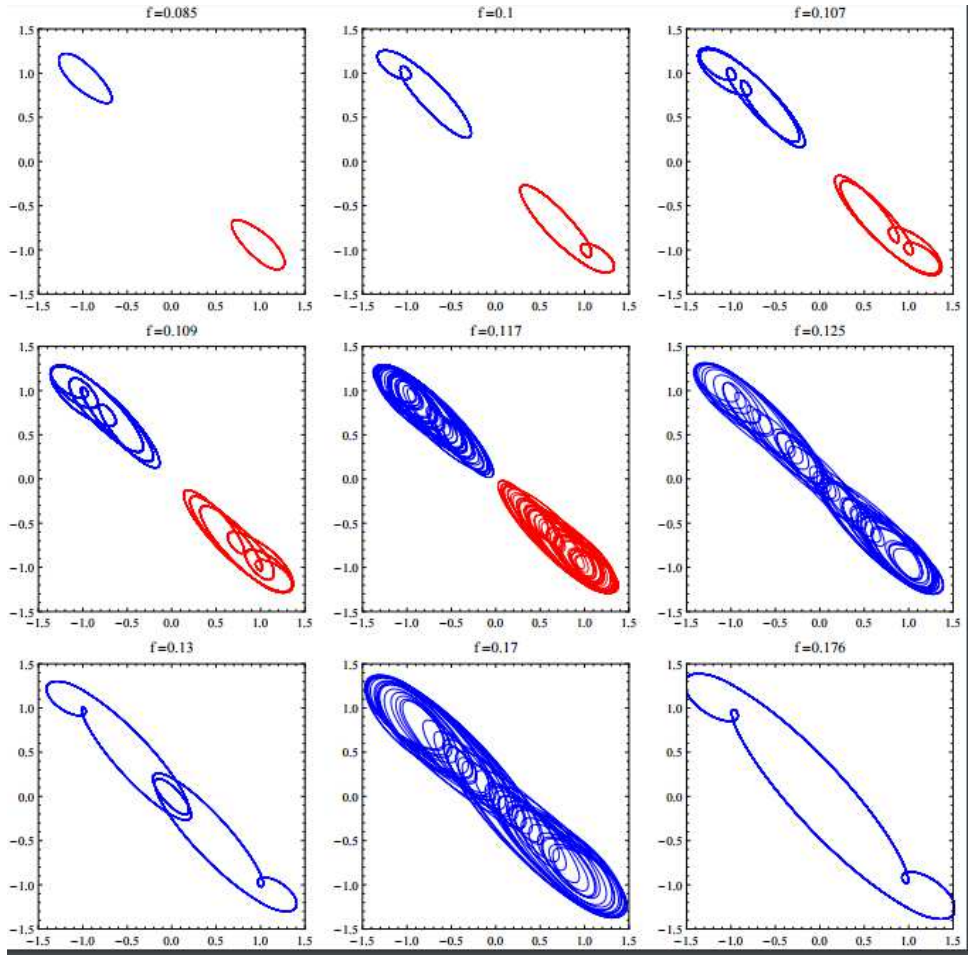}
\caption{Rich dynamic behavior in system (\ref{sys.LRC}), for $\alpha_1=\alpha_2=0.9$ and different values of the parameter $f$.}
\label{fig:per.doubling}
\end{center}
\end{figure}

In the absence of the forcing term (i.e. $f=0$), system (\ref{sys.LRC}) is autonomous and has three equilibrium states: $E_0=(0,0)$ and $E_\pm=\left(\pm\frac{\sigma(a-b)}{1+\sigma b},\pm\frac{b-a}{1+\sigma b}\right)$. However, when $f>0$, the system (\ref{sys.LRC}) is non-autonomous and a series of period-doubling bifurcations leading to onset of chaotic behavior has been reported \cite{Palanivel2017} when $f$ is increased from $0$ to $0.2$, considering the fractional orders $\alpha_1=\alpha_2=0.9$. 

Using the HPC implementation of the parallel algorithm described in section 3, we are able to depict the dynamic behavior of system (\ref{sys.LRC}) with an improved precision compared to \cite{Palanivel2017}, using a small step size and computing the numerical solution over a large time interval.

Figure \ref{fig:per.doubling} shows the attractors of  (\ref{sys.LRC}), for different values of the parameter $f$. For $f=0.085$, the existence of two quasi-periodic attractors is observed and the period-doubling cascade actually involves both attractors, eventually leading to the appearance of two chaotic attractors (e.g. for $f=0.117$). When the value of $f$ is increased, these chaotic attractors collide and a double-scroll attractor takes their place (e.g. for $f=0.125$). As we further increase $f$, a sequence of period-doubling bifurcations and reversed period-doubling bifurcations is observed, involving the single attractor of the system.

\section{Conclusion and future work}

By taking a closer look at the hardware architecture we obtained an improvement on the running time by decreasing the communication time between the process. A similar algorithm was implemented to run on PC using MPI or/and OpenMP with similar results \cite{XingCai,Zhang2014performance}. Additionally, by running our MPI implementation combined with OpenMP on the available BlueGene/P hardware, we gain 10\% running time performance. Moreover, in order to improve the computation running time, some hardware features and capabilities were exploited, leading to a 10 fold reduction of the overall simulation time, with the expense of loosing data precision. 
	
The algorithm that implements the Adams-Bashforth-Moulton method is valid for solving any kind of fractional-order system with fractional derivatives of Caputo-type, hence, having the algorithm run with parameters and functions as input and execute the simulation as a black box is one of the future research objectives.

As another direction for future research, other numerical methods, possibly using nested meshes, for solving fractional-order systems of ordinary differential equations or partial differential equations \cite{gong2014efficient} will be explored, as well as their parallel implementation algorithms. 
 
\section*{Acknowledgements}

This work was supported by a grant of the Romanian National Authority for Scientific Research and Innovation, CNCS-UEFISCDI, project no. PN-II-RU-TE-2014-4-0270.

\bibliographystyle{spmpsci}

\end{document}